\documentclass[11pt,a4paper]{article}
\usepackage{amsfonts,amsmath,amssymb,amscd,mathrsfs,comment,enumitem,graphicx}
\usepackage[usenames,dvipsnames]{xcolor}
\definecolor{CeruleanBlue}{rgb}{0.16, 0.32, 0.75}

\usepackage{newtxtext,newtxmath}  

\usepackage{graphicx}
\usepackage{bm}
\usepackage{physics}
\usepackage[version=4]{mhchem}
\usepackage{mathtools,mathrsfs,amsfonts,dsfont}
\usepackage{relsize}
\usepackage{scalerel}
\usepackage{todonotes}
\usepackage{ulem}
\usepackage[unicode=true, pdfusetitle, bookmarks=true, bookmarksnumbered=false, bookmarksopen=false, breaklinks=true, pdfborder={0 0 0}, backref=false, colorlinks=true, allcolors=CeruleanBlue]{hyperref}
\usepackage{enumitem}
\usepackage{cancel}
\usepackage{float}
\usepackage{authblk}
\usepackage[labelfont=bf]{caption}
\usepackage{subcaption}

\newcommand{\sm}{\kern0.1em}
\newcommand{\smalldiv}{\raisebox{-0.2ex}{\resizebox{!}{1.6ex}{\sm/\sm}}}
\newcommand{\tf}{t_{\!f}}

\usepackage[margin=3cm]{geometry}

\usepackage[
backend=biber, 
style=phys, 
eprint=false,
doi=false,
sorting=none,  
biblabel=brackets,
bibencoding=utf8]{biblatex}
\begin{document}

\title{Double-slit experiment revisited}

\author[*]{Siddhant Das}
\author[*]{Dirk-Andr\'e Deckert}
\author[*]{Leopold Kellers}
\author[$\dagger$,$\ddag$]{Simon Krekels}
\author[$\dagger$,$\$$]{Ward Struyve}
\affil[*]{Mathematisches Institut, Ludwig-Maximilians-Universit\"{a}t M\"{u}nchen}
\affil[$\dagger$]{Department of Physics and Astronomy, KU Leuven}
\affil[$\ddag$]{Imec, Leuven}
\affil[$\$$]{Centre for Logic and Philosophy of Science, KU Leuven}

\maketitle
\normalem

\begin{abstract}
The double-slit experiment is one of the quintessential quantum experiments. However, it tends to be overlooked that a theoretical account of this experiment requires the specification of the joint position and time distribution of detection at the screen, whose position marginal yields the famous interference pattern. The difficulty then arises what this distribution should be. While there exists a variety of proposals for a quantum mechanical time observable, there is no consensus about the right choice. Here, we consider Bohmian mechanics, which allows for a natural and practical approach to this problem. We simulate this distribution in the case of an initial Gaussian wave packet passing through a double-slit potential. We also consider a more challenging setup in which one of the slits is shut during flight. To experimentally probe the quantum nature of the time distribution, a sufficient longitudinal spread of the initial wave packet is required, which has not been achieved so far. Without sufficient spread, the temporal aspect of the distribution can be treated classically. We illustrate this for the case of the double-slit experiment with helium atoms by Kurtsiefer {\em et al.}\ \href{https://www.nature.com/articles/386150a0}{Nature \textbf{386}, 150 (1997)}, which reports the joint position and time distribution. 
\end{abstract}

\section{Introduction}
Diffraction and interference phenomena occupy a prominent place in the phenomenology of quantum physics, Young's double-slit experiment (DSE) being an archetypal example which, according to Feynman ``has in it the heart of quantum mechanics'' \cite{Feynman}. Quantum physicists have discussed the DSE and augments thereof, e.g., the which-way \cite{Zurek,WWM}, or the delayed-choice DSE \cite{Wheeler}, with varying degrees of rigour. Several notable realizations employing \emph{single} electrons \cite{Batelaan,tonomura}, neutrons \cite{neutronsA,neutronsB}, atoms \cite{atomDS,Pfau}, and even macro-molecules \cite{molecules} have been performed. 

The interference fringes in a DSE are formed by the accumulation of single-particle impact positions on a screen. What tends to be overlooked, however, is that each imprinted position with screen coordinates \((x,y)\) is associated with a definite, but random, time of flight\footnote{Also known as arrival time, transit time, or detection time.} (ToF) \(\tf\), given by the (measured) time of detection on the screen minus the (known) time of emission at the source. Therefore, specifying the joint distribution \(\rho(x,y,\tf)\)---whose marginal \(\int\! \dd{\tf} \rho(x,y,\tf)\) provides the pattern of the fringes---is indispensable for a satisfactory account of the DSE, or any other scattering experiment for that matter\cite{werner86,DDGZ,viale,DurrTeufel}.

One might be inclined to think that the distribution $\rho(x,y,\tf)$ on the screen corresponds to $\sm|\psi(x,y,d,\tf)|^2$, where \(\psi(x,y,z,t)\) is the particle's wave function at time \(t\) and \(\smash{z=d}\) the detection plane. But this is incorrect because \(|\psi(x,y,z,t)|^2\) is, according to Born's rule, the distribution of \emph{positions} at a \emph{specified time} \(t\), while what is needed is a joint distribution of \emph{positions} and \emph{times} detected at a \emph{specified surface}.\footnote{The conjectured quantity is not even normalizable in simple examples. Therefore, it cannot be a legitimate probability density.} 

Nevertheless, under certain conditions, such as a narrow spread in the longitudinal direction, i.e., in the direction towards the screen,
which tend to apply to present-day experiments, the detection times at the screen can be treated classically, leading in effect to a joint distribution that is well approximated by $|\psi|^2$ \cite[Eq. (9)]{viale}. This is something that needs to be taken into account in order to really probe the quantum nature of the time distribution.

While there is no recognized observable for time in quantum mechanics, in contrast to position, there is a multitude of (disparate) ToF proposals \cite{MUGA1,MSP,backwall1} (some of them being questionable \cite{Mielnik,LeavOTS,Mugareply,Leavensreply,gaugeinv,WardNico,cavendish2024}). Barring a few works checking tunneling-time predictions, e.g., \cite{Spielmann, Steinbergphoton,LandsmannPRL,AttoclockReview,SteinbergTunneling}, surprisingly, so far none of the ToF proposals have been benchmarked against experiment. 

The goal of this paper is to consider the Bohmian approach to this problem, by considering the first-passage (or hitting) distribution, i.e., the joint distribution of times and locations of particles crossing the detection screen. While this approach ignores the possible effect of the screen itself, this seems a natural and practicable approximation. The potential for treating ToF experiments this way has long been recognized \cite{DDGZ,Leav,Leav98,Grubl,cushing}, with numerous applications to arrival- and tunneling-time problems; see \cite{DDSE,Wuhan,Leavens1996,Leavensspeedup,Nicolas,DD,Exotic,Oriolstunneling,Mousavi,Mousavitunnel,Nogami,BohmbeatsKij,Home,CPviol,HomeCP,Golshani,Demir,TunnelingTime}. In section \ref{sec:bohmian}, we will consider the DSE setup and simulate the evolution of an initial Gaussian wave packet passing through the slits. The first-passage distribution of the Bohmian trajectories is considered for three different placements of the detection screen, leading to qualitatively different behavior. Subsequently, we consider a challenging variant of the DSE, dubbed the dynamic DSE, in which one of the slits is shut in flight. This serves as an illustration of the potency of the adopted approach. 

\begin{figure}[!ht]
	\centering
	\includegraphics[width=\columnwidth]{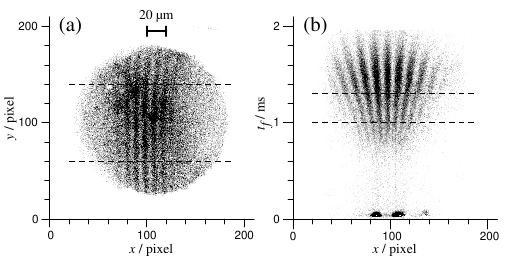}
	\caption{Figures from the KPM experiment \cite{Pfaudetector} (reproduced with permission). \textbf{(a)} Double-slit interference pattern observed on a screen after many single-atom detection events. \textbf{(b)} Joint distribution of the ToF \(\tf\) and the \(x\)-screen-coordinate of the detection events. Impact positions of atoms arriving in the \(\tf\) range indicated by the dashed lines in (b) were accumulated to produce (a). A fraction of very fast atoms emanating from the source cast a shadow of the slits at the bottom.}
	\label{fig:DSE_detections}
\end{figure}

On the experimental side, Kurtsiefer, Pfau, and Mlynek (hereafter KPM) \cite{Pfau,Pfaudetector,wig} reported the joint position distribution using metastable (i.e., electronically excited) helium atoms (see below for further details). In \cite{Pfaudetector}, two marginals distributions are plotted: Fig.~\ref{fig:DSE_detections}a shows the distribution of $(x,y)$ on the screen, while Fig.~\ref{fig:DSE_detections}b shows that of $(x,\tf)$. In \cite{Pfau}, the joint distribution is plotted for three different distances of the detection screen, see Fig.\ \ref{eq:kurtsiefer_hist}. The Bohmian first-passage distribution, shown in \ref{fig:triple_hist}, shows striking similarity. However, this similarity is deceptive, as these plots concern different quantum ensembles. As we will explain in section~\ref{sec:KPM}, in our Bohmian analysis we assumed an ensemble where each wave function is the same and has an appreciable longitudinal spread. The ensemble considered by KPM, on the other hand, concerns a mixture of wave functions whose longitudinal velocity is thermally distributed, and which do not have appreciable longitudinal spread. So in the latter case the temporal spread is not due to the longitudinal spread of the wave function, but due to the random initial longitudinal velocity, allowing KPM to treat the ToF classically. This also seems to be the case in other interference experiments to date. To actually probe the quantum nature of the time distribution, it is important to consider similar experimental setups for which the longitudinal spread of the initial wave function of the mixture has a considerable contribution to the temporal spread of the joint distribution.

\begin{figure}[!ht]
  \centering
    \includegraphics[width=0.8\linewidth]{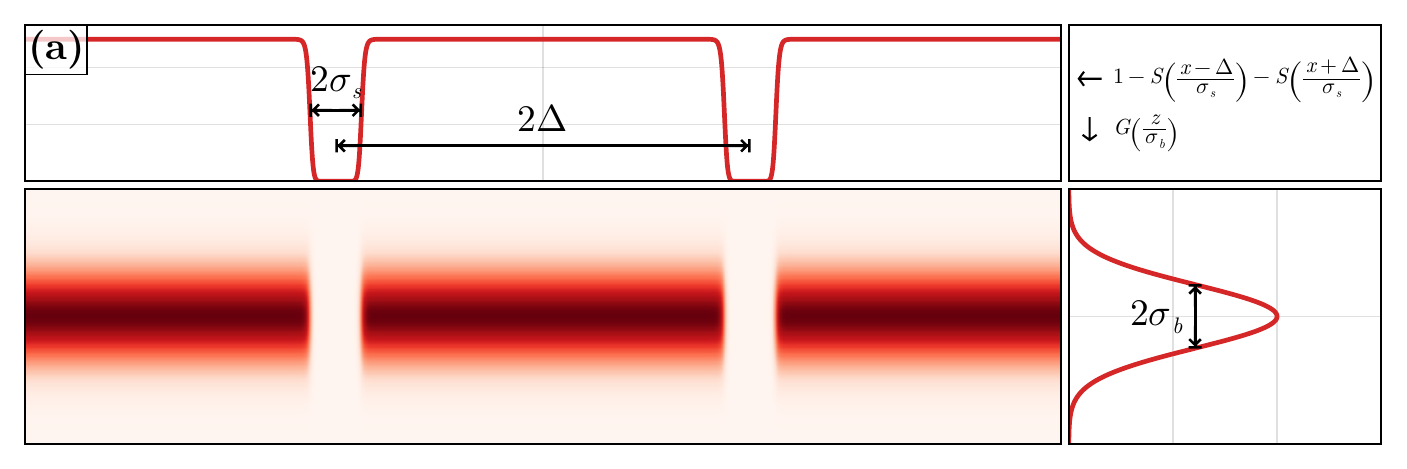}\\
    \includegraphics[width=0.8\linewidth]{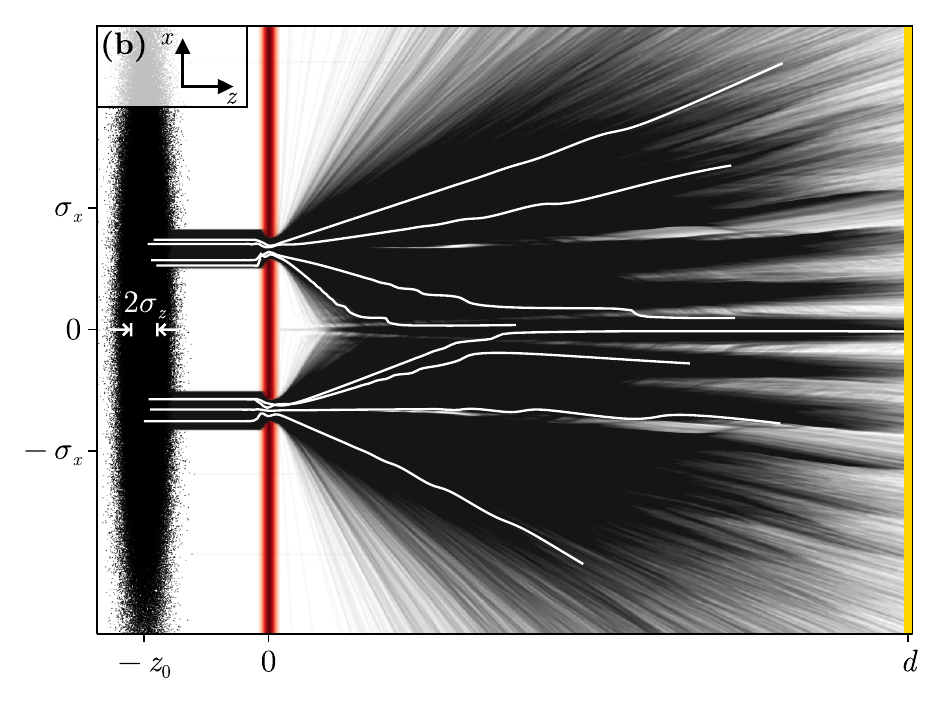}
    \caption{\textbf{(a)} Schematic illustration of the double-slit potential \eqref{eq:pot} for \(\smash{f_1=f_2=1}\). 
    \textbf{(b)} A collection of \(5\times10^5\) Bohmian trajectories in the $xz$-plane for the DSE with initial conditions (black dots) sampled randomly from the \(|\psi(\vb{r},0)|^2\)-distribution. Only the trajectories which pass through the slits (i.e. those which are not back-scattered by the double-slit potential) are shown. Each curve tracks a particle that eventually strikes the detection plane \(\smash{z=d}\). The following potential barrier and wave packet parameters were assumed: \(\smash{V_0=10^3}\), \(\smash{\sigma_b=0.125}\), \(\smash{\sigma_s=0.25}\), \(\smash{k_x=x_0=0}\), \(\smash{k_z=15}\), \(\smash{z_0=2}\), and \(\smash{\sigma_x=3,\sm\sigma_z=0.25}\) in units where \(\smash{\hbar=m=\Delta=1}\). A number of selected trajectories (in white) display the characteristic non-Newtonian meandering of the trajectories.}
    \label{fig:DSE_trajectories}
\end{figure}

\section{Bohmian account of the DSE} \label{sec:bohmian}
\subsection{Trajectories for the DSE}
Bohmian mechanics (also called the de Broglie-Bohm theory or pilot-wave theory) is a theory describing point particles moving along trajectories that grounds the formalism and predictions of standard quantum mechanics \cite{BohmHiley,HollandBook,DurrTeufel,DGZOperators}. For a single-particle with position \(\vb{R}\) and mass \(m\), the dynamics is given by the guidance equation
\begin{equation}\label{eq:guide}
    \dot{\vb{R}}(t)=\frac{\hbar}{m}\sm\Im\!\left[\!\frac{\pmb{\nabla}\psi\big(\vb{R}(t),t\big)}{\psi\big(\vb{R}(t),t\big)}\!\right]\!,
\end{equation}
where \(\psi(\vb{r},t)\) is the wave function satisfying Schr\"odinger's equation with an external potential \(V(\vb{r},t)\). The wave function never collapses in this theory. Here, the ``wave/particle duality'' of quantum mechanics is resolved in a trivial way: both a wave \emph{and} a particle are present. 

The motion of the particle is deterministic, i.e., given its initial position
\(\vb{R}(0)\) and wave function \(\psi(\vb{r},0)\), it has a unique trajectory
$t\mapsto \vb{R}(t)$. However, considering different experimental runs, the initial
positions are random, with distribution given by
\(|\psi(\vb{r},0)|^2\)---the quantum equilibrium distribution---(which implies that the position is distributed according to \(|\psi(\vb{r},t)|^2\) at time $t$). As such, both the detection locations and times of a particle are random in different runs of a scattering experiment. 

Bohmian trajectories for the DSE are shown in Fig.\ \ref{fig:DSE_trajectories}b, and are well-known; they feature invariably in expositions of Bohmian mechanics. DSE trajectories were first presented by Philippidis \emph{et al.}\ in the late 1970s \cite{Dewdney} and have been reproduced various times using different methods \cite{BohmDS,Sanz,holland2,gondran}. Most presentations make do with a freely propagating wave function, not employing a double-slit potential. Here, they are produced without this simplifying assumption. See also \cite{Kocsis} for a weak measurement of average trajectories in a DSE, which resemble those of Fig.\ \ref{fig:DSE_trajectories}b. 

The trajectories provide an almost self-explanatory account of how the interference pattern builds up on a distant screen one particle at a time. In particular, they show ``how the motion of a particle, passing through just one of two holes in [a] screen, could be influenced by waves propagating through both holes. And, so influenced that the particle does not go where the waves cancel out, but is attracted to where they cooperate'' \cite[p.\ 191]{Bell}. It follows that interference experiments can be accounted for in terms of particle trajectories, claims to the contrary notwithstanding.\footnote{For instance, ``it is clear that [the DSE] can in no way be reconciled with the idea that electrons move in paths. ... In quantum mechanics there is no such concept as the path of a particle.'' \cite[p.\ 2]{LandauLifshitz}, or that ``many ideas have been concocted to try to explain the curve for \(P_{12}\) [the
interference pattern] in terms of individual electrons going around in complicated ways through the holes. None of them has succeeded.'' \cite[Sec.\ 1.5]{Feynman}.} 

To produce the Bohmian trajectories in Fig.\ \ref{fig:DSE_trajectories}b, Schr\"odinger's equation was solved numerically with initial Gaussian wave function centered at \((-\sm x_0,0,-\sm z_0)\) in front of the slits:
\begin{equation}\label{eq:GGG}
    \psi(\vb{r},0)= \frac{e^{i(k_xx+k_zz)}}{\sqrt{\sigma_x\sm\sigma_y\sm\sigma_z}} G\!\left(\!\frac{x+x_0}{\sigma_x}\!\right)\kern-0.1em G\!\left(\!\frac{y}{\sigma_y}\!\right)\kern-0.1em G\!\left(\!\frac{z+z_0}{\sigma_z}\!\right)\!,
\end{equation}
where \(\smash{G(\xi)=\pi^{-1/4}\exp(-\,\xi^2/2)}\).

The screen containing the slits was modelled by the Gaussian potential barrier
\begin{align}\label{eq:pot}
    V = V_0\sm G\!\left(\!\frac{z}{\sigma_b}\!\right)\!\left[1-f_1S\!\left(\!\frac{x-\Delta}{\sigma_s}\!\right)-f_2\sm S\!\left(\!\frac{x+\Delta}{\sigma_s}\!\right)\right],
\end{align}
featuring two rectangular apertures for \(\smash{f_1=f_2=1}\),\footnote{For the dynamic DSE problem considered below, \(f_{1,2}\) are varied in time as per Eq.\ \eqref{eq:dynamicT}.} as shown in Fig.\ \ref{fig:DSE_trajectories}a. Here, \(\smash{S(\xi)=1\smalldiv(1+\xi^{16})}\), \(2\Delta\), \(\sigma_s\), and \(\sigma_b\) respectively denote the aperture function, inter-slit separation, slit width, and the thickness of the barrier. Throughout the present work, masses, lengths and times are expressed in units of \(m\), \(\Delta\) and \(\smash{m\Delta^2\kern-0.1em\smalldiv\hbar}\), respectively. (This is equivalent to setting \(\smash{\hbar=m=\Delta=1}\).) The Bohmian trajectories were computed by numerically integrating the guidance equation \eqref{eq:guide} for \(5\times 10^5\) initial positions randomly sampled from the $|\psi(\vb{r},0)|^2$-distribution. About 15\% of the sampled trajectories made it through the slits giving rise to detection events. The rest (not shown in Fig.\ \ref{fig:DSE_trajectories}b) were back-scattered from the potential barrier. 

\subsection{First-passage distribution}
The preceding discussion suggests that it is natural to take, as an approximation to the ToF on a specified surface \(\mathcal{D}\) (such as the plane \(\smash{z=d}\)), the first-passage time (FPT) of a Bohmian trajectory\footnote{One considers the FPT because in general a trajectory might cross the surface \(\mathcal{D}\) more than once \cite{DD,Exotic}. However, this does not happen in the present setting.}:
\begin{equation}\label{eq:def_tf}
    \tf(\vb{r}_0)=\text{inf}\sm\big\{\,t:\vb{R}(t)\in\mathcal{D}\text{ and }\vb{R}(0)=\vb{r}_0\,\big\},
\end{equation}
\(\vb{R}(\tf)\) being the concomitant crossing position on \(\mathcal{D}\),\footnote{The infimum in \eqref{eq:def_tf} ensures that the FPT of trajectories not intercepting \(\mathcal{D}\), e.g., those back-scattered at the plane containing the slits giving rise to non-detection events, is infinity, since \(\text{inf}\sm\emptyset:=\infty\).} as a function of the random initial condition \(\vb{r}_0\)---quantities that have no respective counterparts in standard quantum theory. 

Barring special circumstances, the distribution of this crossing event \(\big(\vb{R}(\tf),\tf\big)\) is only numerically accessible. When there is no backflow (i.e., when trajectories only cross the surface \(\mathcal{D}\) once), like in the cases considered here, and more generally in the far-field or scattering regime, the distribution reduces to \(j_\perp(\vb{r},\tf)\) \cite{Kreidl,DDGZ,DDGZ96}, \cite[p.\ 7]{gaugeinv}---the component of the quantum flux (or probability current) density orthogonal to the surface \(\mathcal{D}\).\footnote{In one-dimensional settings there is an explicit formula \cite[Eq.\ (12)]{Kreidl} for the distribution of \(\tf\) defined in Eq.\ \eqref{eq:def_tf}.}  

Note that the FPT does not take into account the details of the detection. This is justified if individual detection events are triggered close to the location and time impact of Bohmian trajectories, and if the detection events do not significantly disturb the trajectories. In the present case, these assumptions seem very natural, especially in the far-field regime. 

However, to be clear, it is also the case that these assumptions cannot hold in general \cite{DDGZ,viale,DurrTeufel,Vona1,goldstein24}.{\footnote{One of us, Siddhant Das, does not agree with the statements expressed in this paragraph, see \cite{DA}.}} First, according to Bohmian mechanics (as well as quantum mechanics), the ToF distribution must be given by a POVM \cite[pp.\ 186-187]{tumulka22}, whereas the FPT distribution generically is not a POVM, not even in the case it is given by the flux \(j_\perp(\vb{r},\tf)\) \cite{Vona1}. Second, there are also other Bohmian-type theories, which concern different guidance equations, but which are nevertheless empirically equivalent, like Nelson's stochastic mechanics \cite{nelson66} and the zig-zag dynamics \cite{colin11,struyve12}. Clearly, such theories will disagree on the FPT (see e.g., \cite{NelsonDSE} and \cite{MAES2022127323} for the FPT distribution in the case of the double-slit experiment). Nevertheless, because of their empirical equivalence, they will agree on the ToF. As such, arrival time experiments like the one considered here can not be used to distinguish such versions of quantum mechanics, contrary to what is sometimes claimed, see for instance \cite{Ayatollah}. They can merely benchmark the different approaches to model the ToF distribution.

A rigorous justification of the FPT as an approximation to the ToF would require actually modelling the detector, but this is easier said than done. In any case, it seems pressing to explore the ToF distribution also experimentally. In setups as considered here, an analysis of the FPT can provide valuable pointers for such endeavours.

\subsection{Numerical DSE}
\begin{figure}
    \centering
    \begin{subfigure}[t]{0.49\linewidth}
        \centering
        \includegraphics[width=\linewidth]{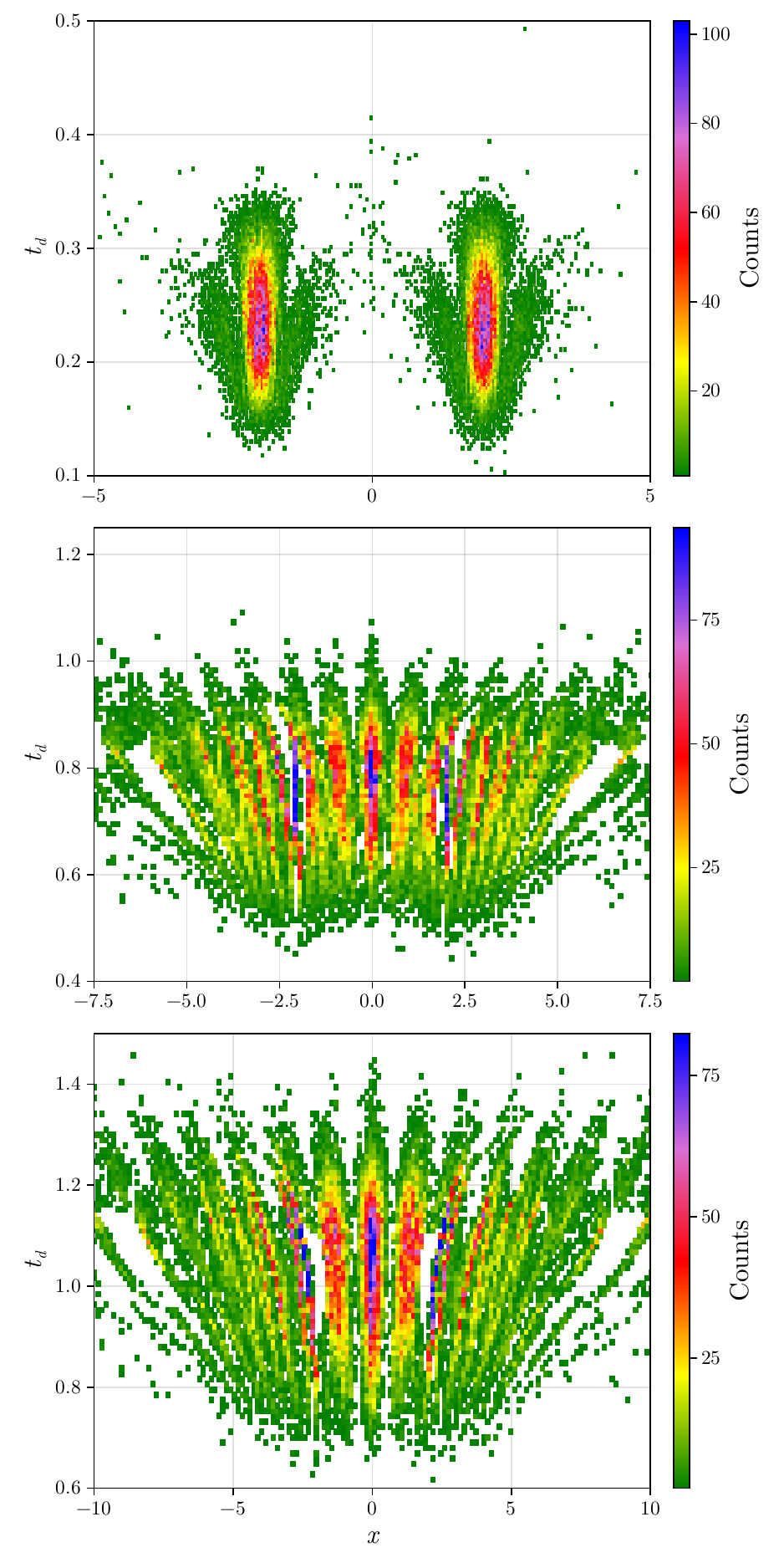}
        \caption{Joint distributions of ToFs and \(x\)-screen coordinates for different slit-screen separations, generated from Bohmian trajectory impacts.}
        \label{fig:triple_hist}
    \end{subfigure}
    \hfill
    \begin{subfigure}[t]{0.49\linewidth}
        \centering
        \includegraphics[width=\linewidth]{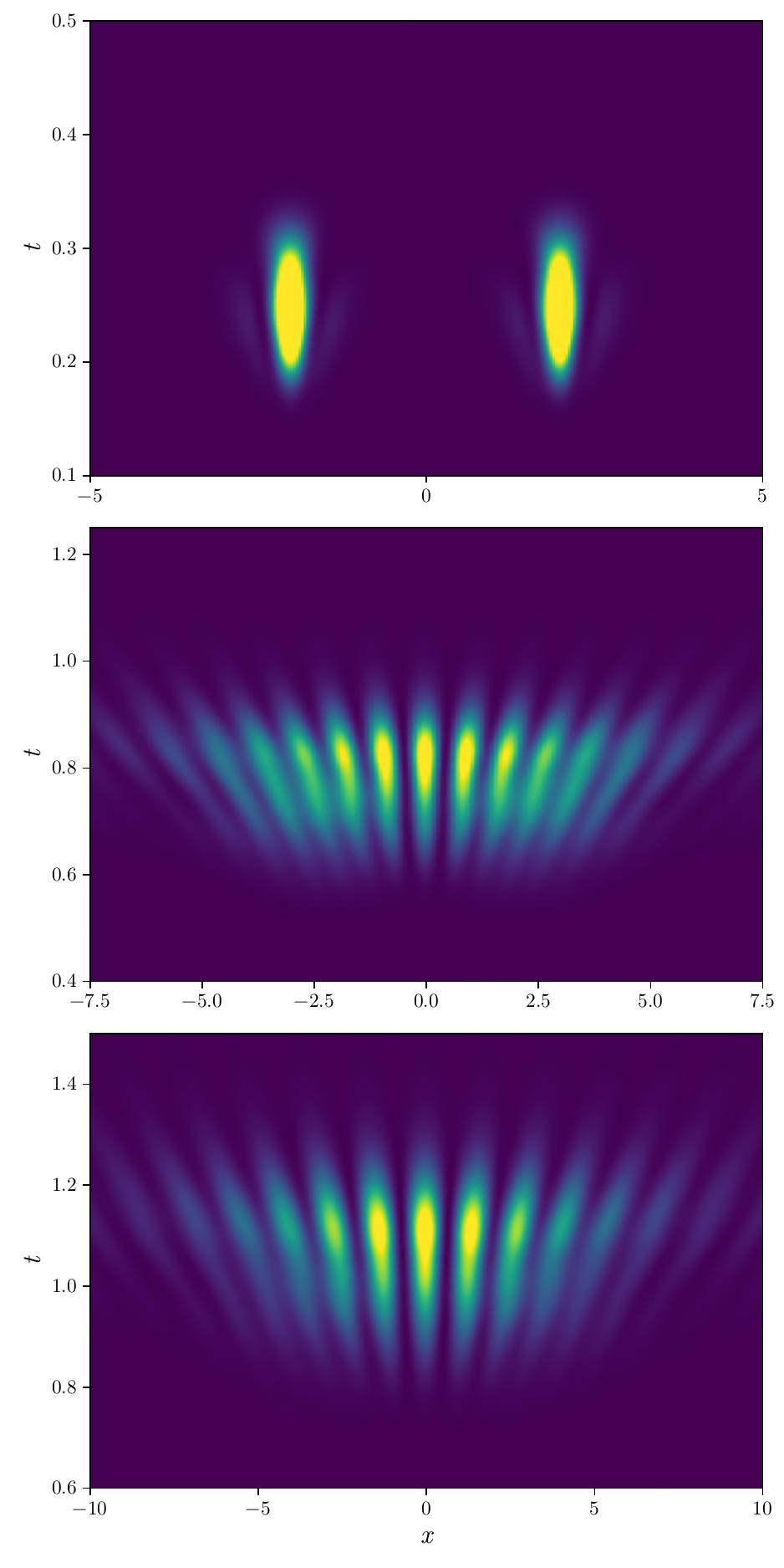}
        \caption{The flux density across the screen.}
        \label{fig:triple_curr}
    \end{subfigure}
    \caption{Numerical data generated for \(\smash{V_0=10^3}\), \(\smash{\sigma_b=0.125}\), \(\smash{\sigma_s=0.5}\), \(\smash{k_x=0}\), \(\smash{k_z=15}\), \(\smash{z_0=3}\), \(\smash{\sigma_x=3}\), \(\smash{\sigma_z=0.25}\), with \(\smash{d=0.9}\) (top), \(\smash{d=10}\) (middle), and \(\smash{d=15}\) (bottom), taking \(\smash{\hbar=m=\Delta=1}\). (The slits were closed after the passage of the wave packet to avoid reflection from the boundary of the grid, see \eqref{eq:dynamicT}; $t_c=0.5$, $\gamma=4$.)\label{bohmianplots}}
\end{figure}

To simulate the first-passage distribution for the DSE, Schr\"odinger's equation with the double-slit potential \eqref{eq:pot} was solved numerically for an initial Gaussian wave packet \eqref{eq:GGG}.
The wave function then determines the particle velocity in the guidance equation \eqref{eq:guide}. 

Since the wave function evolution is simulated on a grid with Dirichlet boundary conditions, we need to avoid effects arising from  reflection at the boundary, in particular reflection from the back wall. To achieve this, the slits were closed dynamically after the passage of the bulk of the wave packet, by changing the $f_{1,2}$ in \eqref{eq:pot} to
\begin{equation}\label{eq:dynamicT}
  f = \frac{1}{2}\sm\bigg(\sm1+\tanh\kern-0.1em\big[\gamma\sm(t_c-t)\big]\sm\bigg).
\end{equation}
This closing of the slits of course also cuts off the back tails of the wave packet, but without significant changes of the FPT distribution of the particles passing through the slits.

Numerically integrating the guidance equation Eq.~\eqref{eq:guide} for \(5\times 10^5\) initial conditions randomly sampled from the \(|\psi(\vb{r},0)|^2\) distribution, the crossing position \(\big(X(\tf),Y(\tf)\big)\) and time \(\tf\) of Bohmian trajectories through \(\smash{z=d}\) were recorded.\footnote{Since the wave function remains a separable function of $(x,z)$ and $y$, the particle motion in the $y$-direction is independent of the motion in the $(x,z)$-plane and hence does not affect the FPT \(\tf\) nor the \(x\)-screen-coordinate \(X(\tf)\).} Histograms of obtained \(\tf\) and \(X(\tf)\) are presented in Fig.\ \ref{fig:triple_hist}, and current densities \(j_\perp(\vb{r},\tf)\) at the screens are shown in Fig.~\ref{fig:triple_curr}.
The code used for the numerics is publicly available \cite{DSEnumerics}.

\begin{figure}
    \centering
    \includegraphics[width=\linewidth]{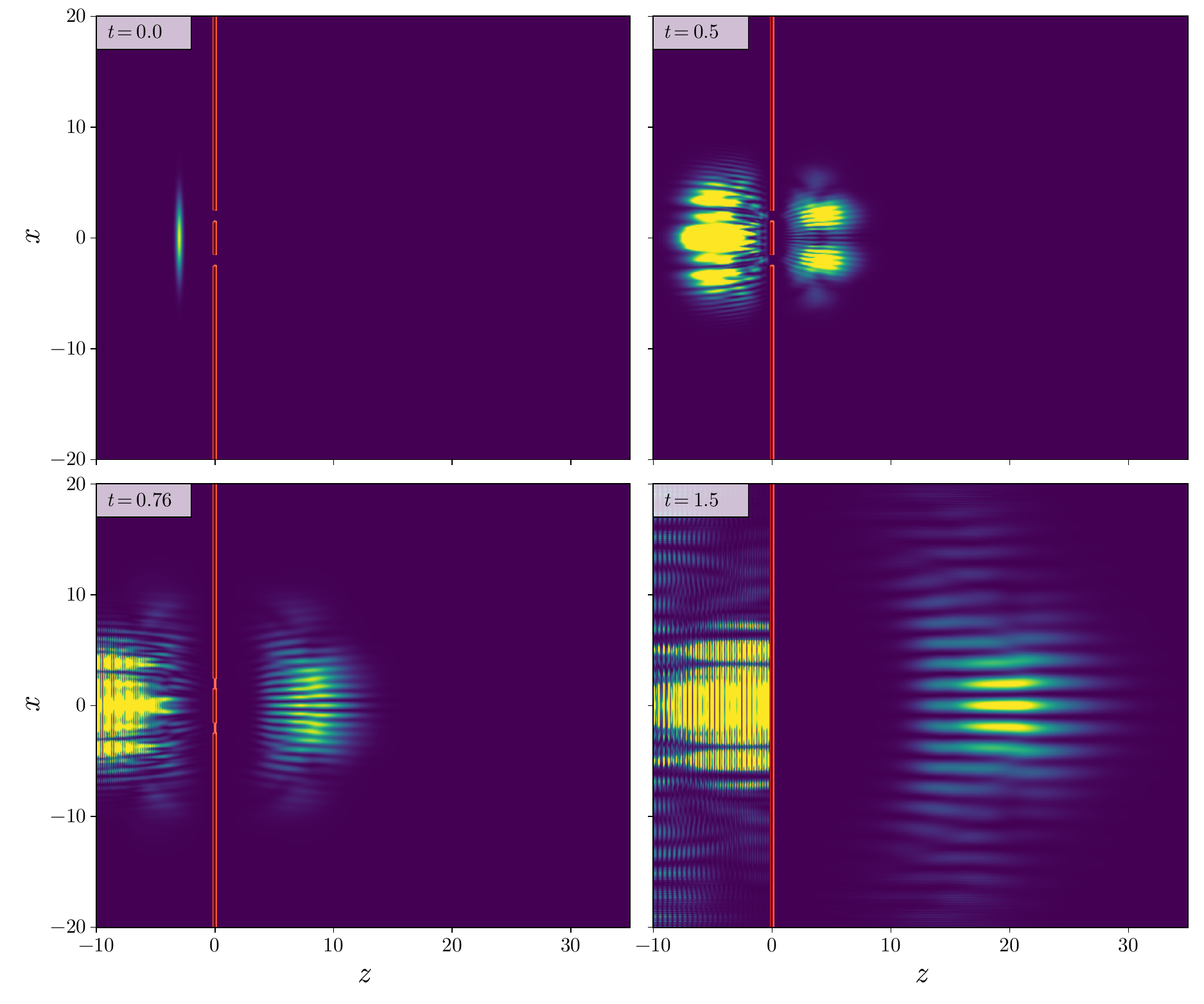}
    \caption{A few snapshots of the evolving $|\psi|^2$-density, using the same parameters as in Fig.~\ref{fig:triple_hist}. Note the closing of the slits (see \eqref{eq:dynamicT}; $t_c=0.75$, $\gamma=4$) to prevent the reflected wave from influencing the arrivals.}
    \label{fig:psi2-example}
\end{figure}

\subsection{Dynamic DSE} \label{dynamic}
It is to be expected that the Bohmian method is also suitable for other setups of this type. To demonstrate this, we consider a more complex time-dependent setup, where one of the slits is shut in flight. For this, we modify only $f_1$ in \eqref{eq:pot}, which causes the slit centred at \(\smash{x=-\sm\Delta}\) to gradually close in time\footnote{A dynamically controlled DSE seems well-amenable to present-day technology \cite{Batelaan}.}, keeping  $f_2=1$. Scatter plots of \(\big(X(\tf),\tf\big)\) for this dynamic DSE are plotted in Fig.~\ref{eq:kurtsiefer_hist} for two values of \(\gamma\); a large \(\gamma\) implies a fast closing of the slit. The closing time \(t_c\) is chosen close to the time at which the peak of the wave packet crosses \(\smash{z=0}\), i.e., \(\smash{z_0\smalldiv k_z}\). Evidently, a smaller number of trajectories pass through the closing slit. Furthermore, the interference fringes become less pronounced for \(\smash{\tf\gg t_c}\), as would be expected for single-slit diffraction.

Note that the fast closing of the slits sends a shock wave through the wave packet, resulting in the banded arrivals at \(t\approx0.6\), visible in both (a) and (b). 
In addition, in the case of a slow closing slit, the arrivals end earlier. This is because for small $\gamma$, while the closing of the slit goes slower, it also starts to close earlier.

\begin{figure}
    \centering
    \includegraphics[width=\columnwidth]{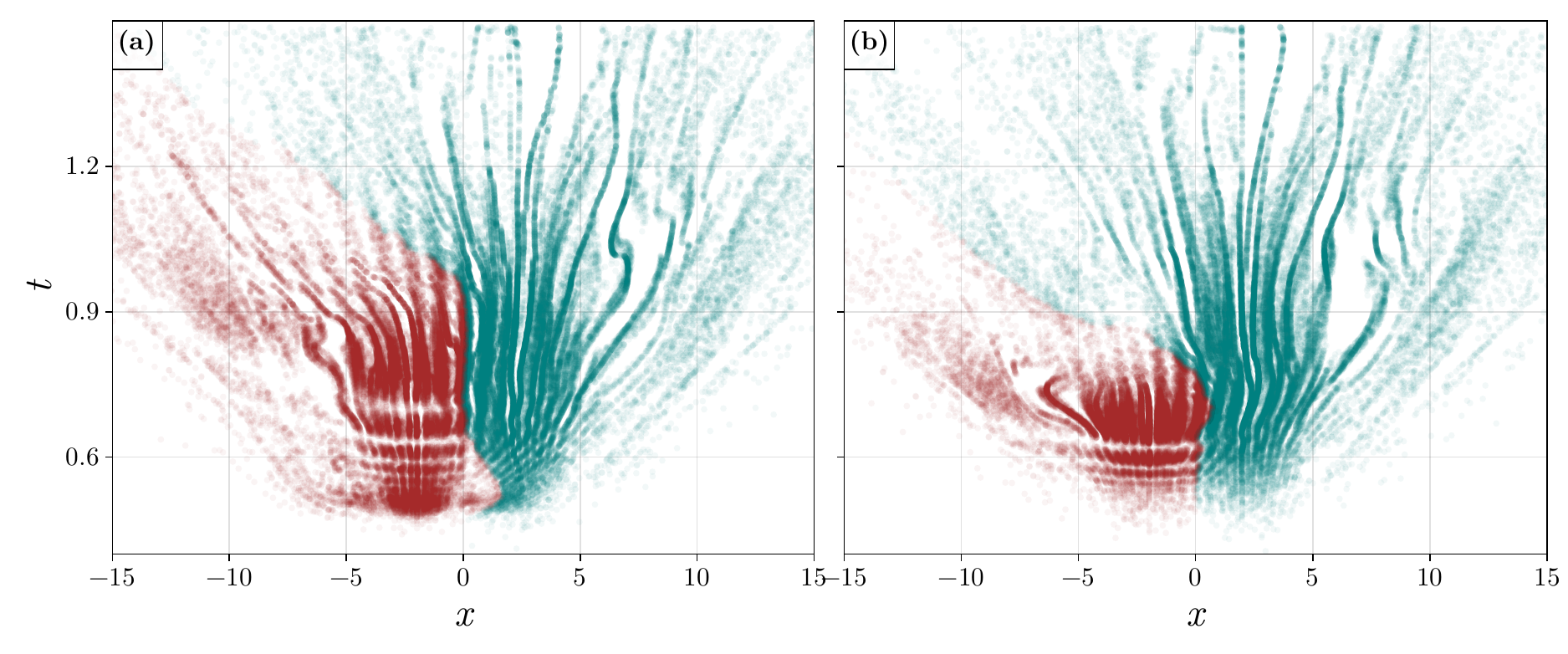}
    \caption{Scatter plot of \(\smash{\approx10^5}\) Bohmian trajectories arriving at \(\smash{d=10}\) over time for the dynamic DSE, where the slit centred at \(\smash{x=-\sm\Delta}\) (\(\smash{=-\sm1}\) in our units) is closed in flight around time \(\smash{t_c=0.25}\): \textbf{(a)} Fast switching \(\smash{\gamma=100}\), \textbf{(b)} Slow switching \(\smash{\gamma=20}\). Data points for trajectories passing through the closing (open) slit are rendered in brown (cyan). The potential barrier and initial wave packet parameters are same as for Fig.\ \ref{fig:triple_hist}.}    
    \label{fig:closing_slit}
\end{figure}

\section{Comparison to the KPM experiment} \label{sec:KPM}
\begin{figure}
    \centering
      \includegraphics[width=0.4\textwidth]{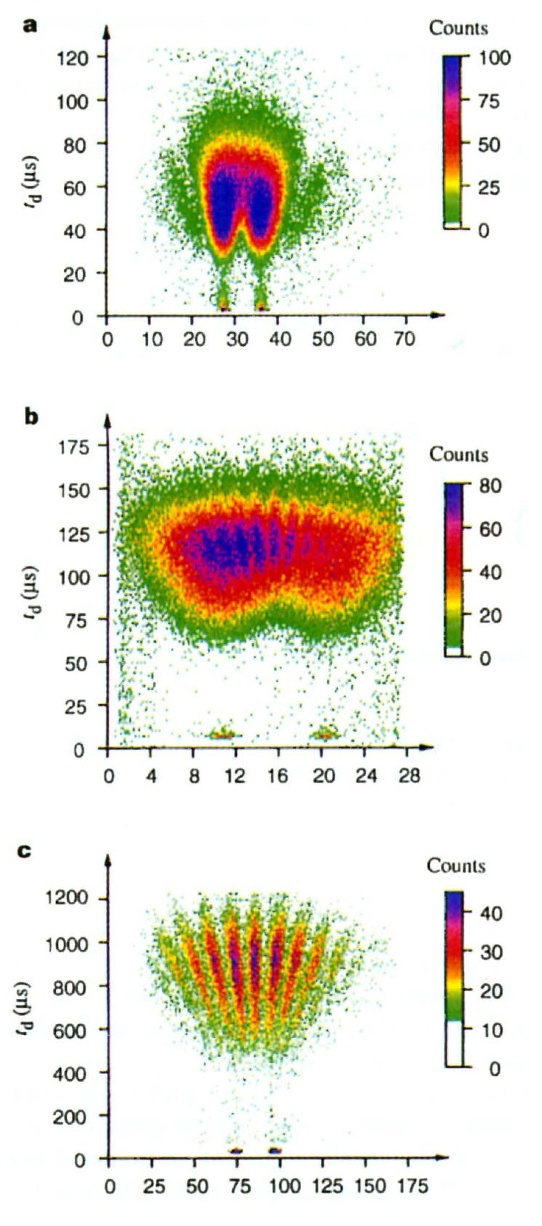}
       \caption{Experimental data for \ce{He^*} atoms with \(\smash{d=148\,\text{mm}}\) (a), \(\smash{d=248\,\text{mm}}\) (b), and \(\smash{d=1,950\,\text{mm}}\) (c) (reproduced from \cite{Pfau} with permission). Here, \(\smash{t_d=(d/L)\,\tf}\) is a ``scaled time-of-flight'', where \(L\) is the source-to-screen separation. }
    \label{eq:kurtsiefer_hist}
\end{figure}

As mentioned in the introduction, there have been many realizations of the double-slit experiment. One experiment of particular interest is that of KPM \cite{Pfau}, which recorded not only the detected positions, but also the flight times. The resulting joint position and time distribution is displayed in Fig.\ \ref{eq:kurtsiefer_hist}, for three different locations of the detection screen. These experimentally obtained distributions show qualitative similarity with the Bohmian distributions shown in Fig.\ \ref{fig:triple_hist}. However, as we will explain here, this similarity is deceiving. 

Let us first consider the details of the KPM experiment. In the experiment, metastable helium atoms (denoted \ce{He^*}) are sourced from a gas-discharge tube by firing a \(15\,\mu\text{s}\) pulse of electric current. A few spurious fast-moving atoms along with slower atoms with velocities ranging between \(\smash{1-3\times10^3\,\text{ms}^{-1}}\)  are produced by the source. The ejected atoms pass through a skimmer and are collimated by a \(\smash{5\,\mu\text{m}}\)-wide slit. Further downstream, they encounter a microfabricated double-slit structure of inter-slit separation \(\smash{2\Delta=8\,\mu\text{m}}\) and slit-width \(\sigma_s=1\,\mu\text{m}\), eventually striking the detector plate at \(d\). \ce{He^*} atoms eject secondary electrons upon hitting the conductive surface of the detector---a process assumed to take place on a time scale of \(10^{-12}\,\text{s}\). These electrons are carefully imaged onto a single-electron detection unit based on a chevron assembly of multichannel plates \cite{Pfaudetector}. The spatial (temporal) resolution achieved with this detector is on the order of \(500\,\text{nm}\) (\(100\,\text{ns}\)). For three different slit-screen separations \(d\), KPM recorded the detection events \((x,y,\tf)\) of individual atoms by repeatedly firing the source, shown in Fig.\ \ref{eq:kurtsiefer_hist}. Notice that the fast atoms produce a ``geometrical shadow'' of the slits at the bottom of each figure \cite[p.\ 151]{Pfau}. In the KPM experiment the ToF from the source to the detection is rescaled as $t_d = (d /L)\tf$, taken to be the time from the passage through the slits to the screen. 

While the KPM plots show a striking similarity with the Bohmian plots, there is an important difference. We have employed {\em the same initial wave function} for each run of the DSE, whereas the discharge tube in the KPM experiment yields a {\em statistical mixture of wave packets} of different longitudinal velocities, thermally distributed, see \cite[Fig.\ 4]{wig}. The mixed state character together with the fact that the longitudinal velocities are a few thousand meters per second implies that the temporal spread of the distributions plotted in Fig.\ \ref{eq:kurtsiefer_hist} stems mainly from the spread of the velocity distribution and {\em not} from the longitudinal spread of the initial wave function. This is also why KPM treat the longitudinal motion as classical. (And why it makes sense to consider the rescaled time $t_d = (d /L)\tf$ to represent the time between detection and passage through the screen.) This is unlike our Bohmian simulations, where the temporal spread in the distribution stems solely from the longitudinal spread of the wave function. 

Let us back up these statements quantitatively. For this, assume that, on the contrary, the temporal spread is largely due to the longitudinal spread of the wave functions and that the mixed state character (which is unavoidable) only plays a minor role. Consider for example the parameters for Fig.\ \ref{eq:kurtsiefer_hist} (c). The temporal spread in this figure is several hundreds of ${\mu}$s. Let us assume therefore that a single wave function can produce a spread in the detection time of at least, say, 100${\mu}$s. Taking a velocity of $10^3$ms$^{-1}$, this means that the longitudinal width of the wave function is of the order of $10^5{\mu}$m. (This is assuming that detections take place while the support of the wave function crosses the detection screen.) 
Assuming Gaussian spreading so that the width $\sigma(t)$ at time $t$ is given by $\sigma(t) =\sigma \sqrt{1+(\hbar t/ m \sigma^2)^2}$, with $\sigma$ the initial width, $m=6.64 \times 10^{-27}$kg the mass of a Helium atom, and taking $t$ to be the time to traverse 3m, which is the distance between the collimating slit and the detection screen (see \cite[Fig.\ 2]{Pfau}), with a velocity of $10^3$ms$^{-1}$, we find that a final width of $\sigma(t)=10^5{\mu}$m implies an initial width $\sigma \approx 5\times 10^{-4}{\mu}$m. This is much smaller than the transverse dimensions of the setup, like the transverse width of $5\mu$m of the collimating slit or the distance between the slits of $8\mu$m. Conversely, assume a longitudinal width of a wave function of $5\mu$m at the time of passage through the collimating slit (so the same as the transverse width of this slit). Assuming again Gaussian spreading, the width will be less than $10\mu$m by the time the wave function crosses the screen. This will entail a temporal spread of merely $10^{-2}\mu$s, which is even smaller than the detector resolution of $10^{-1}\mu$s. 

It is also the case that the temporal spread displayed in the figure can be accounted for by the spread of the velocities. With velocities ranging from $10^3$m to $3\times10^3$m, the associated temporal spread is of the order of $10^3\mu$m, which agrees with what is shown in Fig.\ \ref{eq:kurtsiefer_hist} (c). 

In conclusion, in the KPM experiment the ToF is determined by the time each wave function crosses the detection screen and does not require a quantum treatment (even though this could of course be covered using the Bohmian dynamics). The same holds for other experimental realizations to date. To explore the quantum aspect of the ToF in the case of the double-slit experiment, the wave function should have a ratio of longitudinal spread and velocity that is much larger than the temporal detector resolution. In addition, the initial velocity should be controlled as precisely as possible to reduce velocity spread. Ultracold neutrons which have a velocity of about 1m/s may be a possibility \cite{viale}. Alternatively, one could perhaps consider a different placement of the detection screen, as recently proposed in \cite{Ayatollah}.

\section{Concluding remarks}
The computation of the probability distribution of ToFs of quantum particles is one of the last areas where experts disagree about what quantum mechanics should predict. While a cornucopia of different ToF distributions exists in the literature, one rarely finds a discussion of the joint distribution of impact positions \emph{and} ToFs studied in this work. In this paper, we have considered a Bohmian approach to this problem. 

While the KPM experiment reported a joint position and time distribution, it is not suitable to test quantum aspects of ToF proposals. However, DSE setups do have the potential to do this. To make progress with this question, we recommend that ToF-resolved DSEs and dynamic variants thereof, as described in this work, should be performed with present-day technology to achieve a \emph{quantitative} experiment-to-theory comparison in the future. A particular improvement would be better control of the initial wave function of the particle, e.g., sourcing the particle from an ion trap post-cooling \cite{FHF}, or using field-emission-tip electron wave packets \`a la \cite{HHBB}, both of which offer better initial wave packet control compared to a gas-discharge source.

\paragraph{Acknowledgements.} J.\ Dziewior, C.\ Kurtsiefer, T.\ Maudlin, M.\ Mukherjee, S.\ Goldstein, R.\ Tumulka, N.\ Zangh\`i, and H.\ Ulbricht are thanked for fruitful discussions, H.\ Weinfurter for suggesting the dynamic DSE, and J.\ M.\ Wilkes for editorial inputs. L.K.\ and D.-A.D.\ acknowledge funding from the Elite Network of Bavaria, through the Junior Research Group ``Interaction Between Light and Matter''. W.S.\ is supported by the Research Foundation Flanders (Fonds Wetenschappelijk Onderzoek, FWO), Grant No.\ G0C3322N. We dedicate this work to the memory of Detlef D\"urr.

\printbibliography
\end{document}